\documentclass[aps,twocolumn,epsfig,graphics,showpacs,floatfix,mathbbm,superscriptaddress,lengthcheck,nopacs]{revtex4}

\usepackage{amsmath,amsfonts,amssymb,amsthm,stmaryrd,graphics,graphicx,epsfig,color,times,bbm,psfrag}

\begin{document}

\bibliographystyle{apsrev}

\newcommand{\tr}{\operatorname{tr}}
\newcommand{\uinvnorm}{|\kern-2pt|\kern-2pt|}
\newcommand{\wt}{\operatorname{wt}}
\newcommand{\spectrum}{\operatorname{sp}}
\newcommand{\erf}{\operatorname{erf}}
\newcommand{\erfc}{\operatorname{erfc}}
\newcommand{\supp}{\operatorname{supp}}
\newcommand{\diam}{\operatorname{diam}}

\bibliographystyle{apsrev}

\newcommand{\me}{\mathrm{e}}
\newcommand{\mi}{\mathrm{i}}
\newcommand{\md}{\mathrm{d}}

\newcommand{\cc}{\mathbb{C}}
\newcommand{\nn}{\mathbb{N}}
\newcommand{\rr}{\mathbb{R}}
\newcommand{\zz}{\mathbb{Z}}
\newcommand{\id}{\mathbb{I}}

\newtheorem{definition}{Definition}
\newtheorem{theorem}{Theorem}
\newtheorem{lemma}{Lemma}
\newtheorem{corollary}{Corollary}
\newtheorem{property}{Property}
\newtheorem{proposition}{Proposition}
\newtheorem{remark}{Remark}
\newtheorem{example}{Example}
\newtheorem{assumption}{Assumption}

\setlength{\parskip}{2pt}

\newcommand{\identity}{\openone}
\newcommand{\be}{\begin{equation*}}
\newcommand{\bea}{\begin{eqnarray*}}
\newcommand{\eea}{\end{eqnarray*}}
\newcommand{\ee}{\end{equation*}}
\newcommand{\bra}[1]{\mbox{$\langle #1 |$}}
\newcommand{\ket}[1]{\mbox{$| #1 \rangle$}}
\newcommand{\braket}[2]{\mbox{$\langle #1  | #2 \rangle$}}
\newcommand{\proj}[1]{\mbox{$|#1\rangle \!\langle #1 |$}}
\newcommand{\ev}[1]{\mbox{$\langle #1 \rangle$}}
\def\sign{\mbox{sgn}}
\def\H{{\cal H}}
\def\C{{\cal C}}
\def\E{{\cal E}}
\def\O{{\cal O}}
\def\B{{\cal B}}
\def\one{\ensuremath{\hbox{$\mathrm I$\kern-.6em$\mathrm 1$}}}
\def\tr{ \mbox{tr}}

\bibliographystyle{unsrt}

\title{General entanglement scaling laws from time evolution}
\author{Jens Eisert}

\affiliation{Blackett Laboratory, Imperial College London, Prince
Consort Road, London SW7 2BW, UK}
\affiliation{Institute for
Mathematical Sciences, Imperial College London, Prince's Gardens,
London SW7 2PE, UK}

\author{Tobias J.\ Osborne}

\affiliation{Department of Mathematics, Royal Holloway University of
London, Egham, Surrey TW20 0EX, UK}

\date\today

\begin{abstract}
We establish a general scaling law for the entanglement of a large
class of ground states and dynamically evolving states of quantum
spin chains: we show that the geometric entropy of a distinguished
block saturates, and hence follows an entanglement-boundary law.
These results apply to any ground state of a gapped model resulting
from dynamics generated by a local hamiltonian, as well as, dually,
to states that are generated via a sudden quench of an interaction
as recently studied in the case of dynamics of quantum phase
transitions. We achieve these results by exploiting ideas from 
quantum information theory and making use of the powerful
tools provided by Lieb-Robinson bounds. We also show that there
exist noncritical fermionic systems and equivalent spin chains
with rapidly decaying interactions whose geometric entropy 
scales logarithmically with block length. Implications for the 
classical simulatability are outlined.
\end{abstract}


\maketitle

At the heart of the intriguing complexity of describing quantum
many-body systems is the entanglement contained in the system's
state: if the state is highly entangled, one needs a large number of
parameters to describe it classically. The scaling of the
\emph{geometric entropy} [1--11]
 --  the degree of entanglement of a
distinguished subsystem with respect to the rest  --  for quantum
many-particle systems, such as those encountered in condensed matter
physics, is the crucial parameter 
which quantifies whether the
state is hard or easy to simulate using density-matrix
renormalisation group methods \cite{Verstraete}.

Recently, motivated partially by questions of simulatability, there
has been a considerable effort to precisely characterise scaling
laws for ground-state entanglement, which we call the \emph{static}
geometric entropy [1--11]
Indeed, substantial progress has been made in answering this
difficult question: earlier conjectures, for which there was only
numerical evidence, could be resolved. For example, it is now known
that for gapped bosonic harmonic systems, such as free field models
\cite{BlackHoles}, the geometric entropy scales like the
\emph{boundary area of a distinguished region}, and not the volume
\cite{Area}. The only precise results available at the current time
pertain to \emph{quasi-free} (or \emph{Gaussian}) bosonic and
fermionic models \cite{Fermionic1d,Wolf,Singleshot} and equivalent
1D spin chains. Apart from integrable systems and matrix-product
state hamiltonians (which satisfy an area law by construction
\cite{MPS}), there is a dearth of results concerning static
geometric entropy for systems as simple as the 1D spin-1 Heisenberg
model. How does the geometric entropy
scale for general interacting systems?

There are also very few results available about the strongly related
case of geometric entropy for \emph{dynamically} evolving states
\cite{Dynamics}.
The dynamic geometric entropy occupies centre stage when trying to
simulate systems which undergo a sudden \emph{quench} of a local
interaction, for example, when a system is in a Mott phase 
when
the hopping is suddenly altered. In the Mott phase the 
geometric
entropy is zero and grows as the system evolves 
\cite{Dynamics}. It is far from obvious how the
geometric entropy should scale as a function of time in these and
similar systems dynamically undergoing a quantum phase transition.

In this Letter we establish the first scaling laws for the geometric
entropy of a general class of quantum states that goes significantly
beyond Gaussian models. On one hand, we will show that if any state
of 1D spins whose geometric entropy satisfies a boundary law (i.e.,
it saturates as a function of $n$, the number of spins) is subjected
to dynamics according to an \emph{arbitrary} 1D local model $K$ for
any constant time $t$ then the dynamic geometric entropy will
continue to satisfy a boundary law, albeit saturating at a larger
constant which depends linearly on $|t|$. On the other hand, when
considering the time evolution generated by the local hamiltonian
$K$, the state that results from this time evolution can be thought
of as the {\it ground state of a gapped hamiltonian}, local or with
rapidly decaying interactions \cite{Dynamical}.
All constituents will
eventually become correlated, but the entanglement built up between
remote parts can be bounded, an intuition that we will cast into a
rigourous form.
Hence, this reasoning is a device that allows us to
establish the result that the \emph{static
as well as the dynamical geometric entropy of a large class of models,
including strongly correlated systems, satisfies a boundary law}.

\begin{figure}[ht]
\center{\includegraphics[scale=0.45]{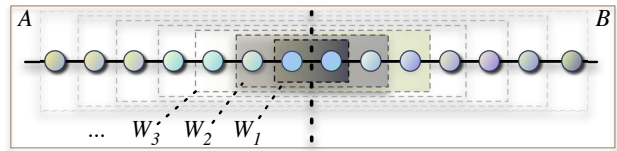}}
 \caption{Hierarchy of unitaries $W_1(t), W_2(t),...$ with exponentially
 decreasing entangling power over
 the boundary of $A$ and $B$
 .}\label{Time}
 \end{figure}

To actually carry out the argument outlined above we use the
powerful machinery of {\it Lieb-Robinson bounds}
\cite{lieb:1972a,OsbSimulation,OsbSimulation2,Osb}. 
The intuition we develop is
that in a many-body system with local interactions there is a {\it
finite speed of sound}, and hence a finite {\it velocity of information
transfer}, resulting from local interactions. The Lieb-Robinson bound
is the precise quantification of this statement: it says that the
norm of the commutator of two operators, one of which is evolving
according to local dynamics, is exponentially small in the
separation between the two operators for short times. This
inequality allows us to precisely bound the entanglement that can
develop across the boundary of a distinguished region for short
times. In turn, we find that for large times of the order of the
logarithm of the number of spins, the boundary law for the dynamic
geometric entropy breaks down. We show this dually by explicitly
constructing a local translation-invariant gapped system whose
ground state violates an area law.

\emph{Geometric entropy in spin chains. -- }We will, for the sake of
clarity, describe our results mainly for a finite chain
$\mathcal{C}$ of $n$ distinguishable spin-$1/2$ particles. The
family $H$ of local hamiltonians we focus on (which implicitly
depends on $n$) is defined by $H = \sum_{j=0}^{n-2} H_j$, where
$H_j$ acts nontrivially only on spins $j$ and $j+1$. We set the
energy scale by assuming that $\|H_j\|$ scales as a constant with
$n$ for all $j = 0,1, \ldots, n-1$, where $\|\cdot\|$ denotes the
operator norm. The interaction terms $H_j$ can, w.l.o.g.,  be taken
to be positive semidefinite, and may depend on {\it time} as $H_j =
H_j(t)$. 

Consider a bi-partition of the chain into two contiguous blocks $A$
and $B$ of spins of sizes $m=|A|$ and $n-m = |B|$, $m<n$. We will
find boundary laws  --  a saturation of the block entanglement  -- 
independent of the system size (we avoid the technicalities arising
in the case of infinite systems which might obscure the main point).
For simplicity we assume $m < n/2$ and we let $|\psi(t)\rangle =
e^{itH}|{0}\rangle$. The initial state is taken to be a product
state $|0\rangle$, but the argument is general enough to be
applicable for any matrix-product or finitely correlated
\cite{Finite,Verstraete} initial state, or a state resulting from a
quantum cellular automata. Consider the Schmidt decomposition
\begin{equation*}
    |\psi(t)\rangle = \sum_{j = 0}^{2^{m}-1}  s_j^{1/2}(t)
    |u_j(t), v_j(t)\rangle,
\end{equation*}
where the $s_j(t)$ are the non-increasingly ordered Schmidt
coefficients. They are given by the eigenvalues of
${M}(t) = {C}(t) {C}^\dag(t)$, where $|\psi(t)\rangle =
\sum_{j=0}^{2^m-1}\sum_{k=0}^{2^{n-m}-1} C_{j,k}(t) |j ,
 k\rangle$, and the $|j\rangle$ and $|k\rangle$ form an
orthonormal basis for $\mathcal{H}_A$  and
$\mathcal{H}_B$, respectively.
The {\it geometric entropy} of a block $A$, or the
\emph{block entanglement}, is given by the von-Neumann
entropy $S(m) = -\sum_{j=0}^{2^m-1} s_j\log_2(s_j)$
\cite{Renyi}.
We denote by $H_A = \sum_{j=0}^{m-2} H_j$ and $H_B =
\sum_{j=m}^{n-2} H_j$ the \emph{local parts} of the
hamiltonian $H$, which act nontrivially only on subsystem $A$ and
$B$, whereas
$H_I= H_{m-1}$ denotes the interaction term. 

{\it Entanglement scaling in dynamically evolving quantum
states. -- }In this section we prove an upper bound for the dynamic
geometric entropy $S(m)$,
    $S(m) \le c_0 + c_1|t|$ ,
where $c_0,c_1>0$ depend only on $\|h\|$ \cite{Norm}
and not on $n$. Thus, the entropy of the
block $A$ scales, asymptotically, less than a constant.
Our first step is to obtain the decomposition
\begin{equation*}
    e^{itH} = (U_A(t)\otimes U_B(t))V(t).
\end{equation*}
We do this by guessing $U_A(t) = e^{itH_A}$ and $U_B(t) =
e^{itH_B}$. The idea here is that the dynamics generated by $H$
should be similar to those generated by $H_A + H_B$ preceded by a
unitary $V(t)$ that ``patches up" the removed interaction.
We obtain a differential equation for $V(t) =
e^{-it(H_A+H_B)}e^{itH} = e^{-it(H - H_I)}e^{itH}$:
$ {dV(t)}/{dt} = V(t)L(t),$.
The ``hamiltonian'' $L(t) = iH_I +\int_{0}^t
\tau_u^{H}([H,H_I])\,du$, where $\tau_t^M(N) =
e^{-itM}Ne^{itM}$ for operators $N,M$, is 
antihermitian, so that the dynamics this
integro-differential equation generates is unitary.

Our strategy at this point is to decompose $V(t)$ into a product of
\emph{strictly} local unitary operations \cite{Sequence}
$V_{\Lambda_{j}}(t)$ which act nontrivially only on $\Lambda_j =
\{x: d(x,m) \le j\}$, which consists of only those sites within a
distance $j$ from the boundary. This decomposition for $V(t)$,
depicted in Fig.\ 1, is then
\begin{equation*}
    V(t) = W_{n-m}(t) W_{n-m-1}(t)\cdots W_2(t) W_1(t),
\end{equation*}
where $W_k(t) = V_{\Lambda_{k}}(t)V_{\Lambda_{k-1}}^\dag(t)$ acts
nontrivially only on $\Lambda_k$.  Also, we set $W_1(t) =
V_{\Lambda_1}(t)$ and $W_{n-m}(t) = V(t) V^\dagger_{n-m-1}(t) $.
Physically, we expect that the unitary operators $W_l(t)$ are
successively weaker and weaker. To find a bound on $\|W_l(t) -
\mathbb{I}\|$ we now invoke the machinery of {\it Lieb-Robinson
bounds} \cite{lieb:1972a} (see Ref.\ \cite{Osb} for a simple direct
proof) on the speed of sound in systems evolving according to local
dynamics: the strongest available such bound \cite{Osb} yields
\begin{equation*}
    \| \tau_t^{H_{\Lambda_l}}(M) -
    \tau_t^{H_{\Lambda_{l-1}}}(M) \| \le
   \delta_l   |t|^l /l! ,
 \end{equation*}
with $\delta_l =  \|M\| 2^l\|h\|^l $, where $M =
[H_{\Lambda_k},H_I]$ is the same for all $k$ as long as $k>1$. This
is indeed the powerful tool we need to derive the desired bound
concerning the deviation from each of the unitaries $W_l(t)$ from
the identity: it exponentially bounds the information spread 
in a system undergoing dynamics under a local hamiltonian.
We hence find
\begin{equation*}\label{eq:wbound}
    \|W_l(t) - \mathbb{I}\| = \|V_{\Lambda_{l}}(t) -
    V_{\Lambda_{l-1}}(t)\| \le
    \delta_l  |t|^{l+2}/(l+2)!.
\end{equation*}
In the last inequality, we have expressed the
operators $V_{\Lambda_{l}}(t)$ as integrals in
time \cite{Technicalities}.
We note that this bound is decaying faster than
exponential in $l$. This bound tells us that we can write
 $W_l(t) = \mathbb{I} +\varepsilon_l(t) X_l(t)$,
where $\|X_l(t)\| = 1$ and $\varepsilon_l(t) \le
\min \{2,  {\delta_l   |t|^{l+2}}/{(l+2)!} \}$.
Let us now consider the action of $e^{itH}$ on the initial product
state vector $|{0}\rangle$,
\begin{equation*}
    e^{itH} |{0}\rangle = e^{it(H-H_I)}
    \prod_{k=1}^{n-m}
    W_k(t)
    |{0}\rangle.
\end{equation*}
We define $|\psi_l(t)\rangle = \prod_{k=1}^{l} W_k(t) |{0}\rangle$.
Let us now choose $l$ large enough so that this bound for $W_l(t)$
is strong enough that $\varepsilon_l(t) \le \min \{2, \|M\|
{\delta_l |t|^{l+2}} / {(l+2)!} \}$ is small \cite{Breakdown}.
Then $|\psi_l(t)\rangle$ is in a product with respect to the
spins outside the region $\Lambda_l$
There are, in general, at most $2^l$ nonzero
Schmidt coefficients for  $|\psi_l(t)\rangle$ with respect
to the bi-partition $AB$.
Now we consider the action of $W_{l+1}(t)$ on $|\psi_l(t)\rangle$
which yields $|\psi_{l+1}(t)\rangle$.
In the computational basis, we have
\begin{equation*}
    |\psi_{l+1}(t)\rangle = \sum_{j,k = 0}^{2^{l+1}-1}
    \bigl((C_l)_{j,k}(t) +
    \varepsilon_{l+1}(t)(D_l)_{j,k}(t)\bigr)
    |j, k\rangle,
\end{equation*}
setting $|\psi_l(t)\rangle = \sum_{j,k = 0}^{2^{l+1}-1}
(C_l)_{j,k}(t)|j, k\rangle$ and $X_{l+1}(t)|\psi_l(t)\rangle =
\sum_{j,k = 0}^{2^{l+1}-1} (D_l)_{j,k}(t)|j , k\rangle$. The
normalisation condition and  $\|X_l(t)\| \le 1$ imply
that  $\|{C}_l\|\le 1$ and $\|{D}_l \|\le 1$. 
%
We use 
Weyl's perturbation theorem \cite{bhatia:1997a} to bound the Schmidt
coefficients $s^{(l+1)}_j$ of $|\psi_{l+1}(t)\rangle$, 
given by the eigenvalues of $(C_l + \varepsilon_{l+1}(t)D_l) (C_l +
\varepsilon_{l+1}(t)D_l)^\dag $.
%
%
%
We apply Weyl's
perturbation theorem to the operators ${P} = C_lC_l^\dag$ and
${Q} =
\varepsilon_{l+1}(t)(C_lD_l^\dag  +
D_lC_l^\dag)  + \varepsilon_{l+1}^2(t)
D_lD_l^\dag$
with $\|{Q}\| \le c\varepsilon_{l+1}(t)$, with $c>0$.
The eigenvalues of
${P}$ are precisely the $2^l$ Schmidt coefficients of
$|\psi_l(t)\rangle$. 
%
%
Weyl's perturbation
theorem tells us that the first $2^l$ eigenvalues of ${P} +
{Q}$ have to be close to the Schmidt coefficients of
$|\psi_l(t)\rangle$ and the remaining $2^l$ eigenvalues have
magnitude less than $\varepsilon_{l+1}(t)$
\cite{Weyl}.
Exploiting these bounds iteratively, we find that
the Schmidt coefficients 
%
%
satisfy the bound
    $s_j(t) \le \min \{{1}/{2^{\kappa|t|}}, 2^{\kappa|t|-v
    j}\}$,
for some $\kappa,v>0$. Hence
the geometric entropy
$S(m)$ satisfies the upper bound
$S(m) \le c_0 + c_1|t|$,
where $c_0,c_1>0$. This holds true for all
$n$. In other words, we can perform
``the limit of infinite system size''
$n\rightarrow\infty$.
When we let $|t| = \log(n)$ our bounds begin to fall apart: the
Lieb-Robinson bound becomes a polynomial bound.
This situation can be saturated, see below.

{\it Entropy-boundary laws for approximately local quantum spin
systems. -- }We now show that the entropy-area law for dynamically
evolving product states implies entropy-area laws for the
\emph{ground states} of noncritical approximately local quantum spin
systems. The product $|{0}\rangle$ is the unique ground state of the
hamiltonian $Z = -\sum_{j=0}^{n-1} \sigma^{3}_j$. Let $H$ be our
hamiltonian. Then $|\psi(t)\rangle$ is the unique ground state of
the new hamiltonian $K = e^{itH}Ze^{-itH}$, having exactly the same
spectrum as $Z$. Moreover, while $K$ is no 
longer strictly local in general, it
is approximately local with exponentially decaying
interactions. The way to see this is to apply a Lieb-Robinson bound
to the interaction term $e^{itH}\sigma_j^3 e^{-itH}$: We consider
the difference between $e^{itH}\sigma_j^3 e^{-itH}$, having support
equal to $\mathcal{C}$, and the strictly local
$e^{itH_{\Lambda_k(j)}}\sigma_j^3 e^{-itH_{\Lambda_k(j)}}$, with
${\Lambda_k(j)} = \{x :  d(x,j)\le k\}$, which has support on $2k+1$
sites. This difference can be bounded using the Lieb-Robinson bound,
    $\|\tau^{H}_t(\sigma_j^3) - \tau^{H_{{\Lambda_k(j)}}}_t(\sigma_j^3)\|
    \le ce^{\kappa|t| - v k}$,
with $c>0$. Thus, the interaction term $\tau^{H}_t(\sigma_j^3)$
couples spins from site $j$ exponentially weakly.
What sort of hamiltonians $K= e^{itH}Ze^{-itH}$ -- clearly
a large class of gapped models --  arise in this way?
Insight can be provided by the following example: Let $H = \sum_{j}
\sigma_j^x\sigma_{j+1}^y$. For small $t$ the hamiltonian $K$ will
look like $K = Z + \lambda(t)\sum_j \sigma_j^x\sigma_{j+1}^x +
\sigma_j^y\sigma_{j+1}^y + O(t^2)$. In this case $K$ is similar to
the $XY$ model in an external magnetic field with small higher
order terms. With the inclusion of 
larger neighborhoods, in turn, local hamiltonians
can be approximated to any accuracy. Another useful 
hamiltonian which can arise in this way
is the strictly local cluster hamiltonian  \cite{Cluster} 
(carrying over also to the higher-dimensional case): 
set $H = \sum_{j}
\sigma_j^x\sigma_{j+1}^x$. In this case, when $t=\pi/2$, $K$ is the
hamiltonian having the cluster state as a unique ground state.

{\it Logarithmic divergence of geometric entropy of gapped
systems. -- }We now construct an explicit situation where a {\it
gapped} 1D spin system ${\cal C}$ 
ndeed violates the entanglement-boundary law. 
We again consider a family of 
spin systems, consisting of a
block $A$ consisting of $m$ spins, and a block $B$ containing the
remaining $n-m$ spins. As before, $S(m)$ is defined 
to be the geometric
entropy of a block $A$ of $m$ spins in the limit of an infinite
chain, for simplicity
with periodic boundary conditions. 
By virtue of the familiar Jordan-Wigner transformation
\cite{JW},
we may  consider the fermionic model
\begin{equation*}
    H= \sum_{l,k=0}^{n-1}
    c_l^\dagger M_{ l-k } c_k,
\end{equation*}
where $M_l\in \rr$, $l=0,\ldots,n-1$. The hermiticity
of $H$ and the periodic boundary conditions are
reflected by the conditions
 $M_l=M_{-l}$ for all $l=0,\ldots,n-1$ and
 $M_l=M_{l+n}$. We can easily map the above
hamiltonian onto the one for non-interacting fermions,
preserving the anti-commutation relations:
    $H= \sum_{k=0}^{n-1}
       \varepsilon_k b_k^\dagger b_k$,
where $\varepsilon_k$, $k=0,\ldots,n-1$, are the eigenvalues of $M$,
given by $\varepsilon_k = \sum_{j=0}^{n-1}
    e^{2\pi i (j+1) k/n} M_j$.
The ground state can then be easily found: it is the
state with unit occupancy for each $k$ with $\varepsilon_k<0$.
If the value $0$ is not contained in the spectrum,
this ground state is non-degenerate.
We now consider the subsystem $A$.
The reduced state $\rho_m$ of this block is
characterised by the spectrum of the real
symmetric $m\times m$ Toeplitz matrix $T_m$
\cite{bhatia:1997a}, which defines the second
moments of fermionic operators
\cite{Fermionic1d,Lieb,Singleshot}.
The $l$-th row of this matrix is given by
$(t_{-l+1},t_{-l+2}, ..., t_0 ,\ldots,t_{m-l})$, where
    $t_l =
    \sum_{k=0}^{n-1} e^{-i l k }
    {\varepsilon_k }/{(n | \varepsilon_k| )}
    $.
At this point, we may take
the limit $n\rightarrow\infty$, for fixed $m$,
and
consider long-ranged interactions, and hence
sequences
of couplings $\{M_l\}_{l\in \nn},\, M_l \in \rr$.
This means that in the continuum limit,
we can consider functions $\phi:(0,2\pi]\rightarrow
\rr$, representing the spectrum of the interaction matrix,
and
    $t_l = {1}/({2\pi})
    \int_{0}^{2\pi}
    e^{-i l x }{\phi(x)}/{|\phi(x)| }  dx$.
We can now make use of a very useful bound of
Ref.\ \cite{Singleshot}, stating that
    $S(m) \geq
    -   (\log_2 |\det[   T_m ] |)/2 $.
Hence, to show that
    $S(m) = \Omega( \log m )  $,
we have to bound the Toeplitz determinant
$\det[T_m]$. This we can do using
a proven instance of the Fisher-Hartwig conjecture \cite{Lieb,FH},
determining the scaling of the determinants of Toeplitz matrices.
Using these ideas, we are
now construct a model with the mentioned surprising
properties: We take the interactions
    $M_k =   \int_0^{2\pi}
    \phi(x) e^{-  i x k}/(2\pi)$ to be given by
       $ M_k  = -i
        (e^{ik\pi/2}-1 )^3 (1+e^{i\pi k/2} ) /(2 e^{2\pi i k}k \pi)$,
so a $1/k$ decay of the interactions, as in case of an
{\it unshielded Coulomb interaction}.
This gives rise to the Fourier transform $\phi$ that takes the
value $1$ in $x \in (0, \pi/2]$, and $(3\pi/2,2\pi]$ and the value
$-1$ in $x\in (\pi/2, 3\pi/2]$. In this setting,
the proven instance of the Fisher-Hartwig
conjecture then indeed allows us to argue that
$   |\det[T_m]| = \Omega(\log m)$ \cite{FH}.
This hamiltonian is obviously {\it gapped}: the quasi-particle
excitation spectrum is even constant, and never crosses zero, so it
defines a gapped system. Still, we find a {\it logarithmically divergent
geometric entropy}. This is an example of a ground state that is not
covered by the above
statement for small times.

{\it Outlook. -- }In this work we have introduced an
 approach to assess geometric entropies in many-body
systems. We have found that many ground states of quasi-local gapped
hamiltonians, while being far from quasi-free, still exhibit a
saturating geometric entanglement, and hence an entanglement-area
law. The studied gapped systems are rigorously classically
efficiently simulatable: one can obtain all expectation values of
local observables with polynomial computational resources
\cite{OsbSimulation}.
Simulatability is closely linked with 1D entropy
boundary laws \cite{Verstraete}. This connection is
even more direct in our
case because matrix-product states which faithfully
approximate our ground states can be {\it explicitly
constructed}
\cite{OsbSimulation2}. Such efficient descriptions in terms of
matrix-product states would also be generated by an
eventually successful application of
the DMRG algorithm to our systems.

Two-dimensional systems are in principle accessible with the methods
introduced here.
This method 
opens up the way toward studying the complexity of 
gapped many-body
systems and the accompanying ground-state entanglement 
scaling (as
well as capacities of quantum channels based on interacting 
systems \cite{Pl}). Intriguingly, we finally found an
example of a gapped system with a divergent block entanglement,
rendering the connection between criticality and validity of an area
theorem more complex than anticipated.

{\it Acknowledgements. -- }We would like to thank M.\ Cramer for
discussions. This work was supported by the
DFG (SPP 1116, SPP
1078), the EU (QAP), the QIP-IRC, 
the Microsoft Research Foundation,
and the EURYI Award of JE.

{\it Note added:} This work complements the simultaneously
submitted Ref.\ \cite{Bravyi}.


\begin{thebibliography}{99}
\bibitem{Dynamical}
    If the initial state
    $|\psi\rangle$ is the unique ground state of some gapped local
    system with hamiltonian $K$ then $e^{itH}|\psi\rangle$ is the
    unique gapped ground state of the system
    $K' = e^{-itH}Ke^{itH}$.
\bibitem{Sequence}
    The sequence of strictly local unitary operations 
    $V_{\Lambda_j}(t)$, $j
    = 1, 2, \ldots, n-m$ is defined by the solutions of
       $  {dV_{\Lambda_j}(t)}/{dt} =
        V_{\Lambda_j}(t)
        (iH_I
        +\int_{0}^t \tau_u^{H_{\Lambda_j}}([H_{\Lambda_j},H_I])\,du
        )$,
    where $H_{\Lambda_j} = \sum_{k\in{\Lambda_j}} H_k$.
    We note that $V_{\Lambda_{n-m}}(t) = V(t)$.
\bibitem{Breakdown}
    Taking the logarithm
    of the quantity ${\delta_l |t|^{l+2}}/{(l+2)!}$ and
    approximating the resulting sum shows us that this
    will happen when
    $l \gg \kappa|t|$, where $\kappa$ is a constant that
    depends only on
    $\|h\|$.
\bibitem{FH}
    More specifically,
    the function $g= \phi/|\phi|$ can be written as
    $\log g(x) = \log a(x) - i \sum_{r=1}^2
    b_r \arg e^{i(x_r + \pi - x)}$,
    where $a$ is sufficiently smooth,
    choosing
    $\arg z \in(0 , 2 \pi ]$. We find the given
    scaling of $|\text{det}[T_m]|$
    \cite{Lieb}
    as now $\beta_{1/2 }= \pm
    1/(2i)$, and hence
    $0<\text{Re} \beta_1<1/2$ and
    $0<\text{Re} \beta_2<1/2$.
\end{thebibliography}
\end{document}